\documentstyle[preprint,aps]{revtex}
\tightenlines
\begin{document}
\draft
\title
{Filling factors and Braid group}
\author
{Wellington da Cruz\footnote{E-mail: wdacruz@fisica.uel.br}} 
\address
{Departamento de F\'{\i}sica,\\
 Universidade Estadual de Londrina, Caixa Postal 6001,\\
Cep 86051-970 Londrina, PR, Brazil\\} 
\date{\today}
\maketitle
\begin{abstract}
We extract the Braid group structure of a recently derived
{\it hierarchy scheme} for the {\it filling factors} 
proposed by us which related the 
{\it Hausdorff dimension}, $h$, to the {\it statistics}, 
$\nu$, of the collective excitations in the context of 
the Fractional Quantum Hall Effect ( FQHE ). 
\end{abstract}

\pacs{PACS numbers: 12.90+b\\
Keywords: Hausdorff dimension; Fractional spin particles;
 Filling factors; Braid group}


In a recent paper\cite{R1}, following ideas established 
in\cite{R2}, we have considered the {\it Hausdorff dimension}, 
$h$, as a parameter which classifies the {\it equivalence classes} 
of the {\it collective excitations} which occurs in the context 
of the FQHE. The elements of each equivalence class are the {\it 
filling factors or statistics}, $\nu$, which characterize the 
collective excitations manifested as quasiholes or 
quasiparticles\cite{R3}. Now, we propose rules of composition 
for these elements such that we extract a {\it Braid group} 
structure from this new hierarchy scheme for the filling factors. 
This scheme takes into account the intervals of definition 
of spin, s, for fractional spin particles, which are related 
to the Hausdorff dimension in such way that we can 
{\it predict} for what values of $\nu$ FQHE can be 
observed\cite{R2}. For a given value of $h$, defined as $1< h < 2$, 
we obtain $\nu$ as ( $i$ means a specific interval ):
        
\begin{eqnarray}
&&h_{1}=2-\nu,\;\;\;\; 0 < \nu < 1;\;\;\;\;\;\;\;\;
 h_{2}=\nu,\;\;\;\;
\;\;\;\;\;\;\;\;\; 1 <\nu < 2;\;\nonumber\\
&&h_{3}=4-\nu,\;\;\;\; 2 < \nu < 3;\;\;\;\;\;\;\;\;
h_{4}=\nu-2,\;\;\;\;\;\;\; 3 < \nu < 4;\;\nonumber\\
&&h_{5}=6-\nu,\;\;\;\; 4 < \nu < 5;\;\;\;\;\;\;\;
h_{6}=\nu-4,\;\;\;\;\;\;\;\; 5 < \nu < 6;\;\nonumber\\
&&h_{7}=8-\nu,\;\;\;\; 6 < \nu < 7;\;\;\;\;\;\;\;
h_{8}=\nu-6,\;\;\;\;\;\;\;\; 7 < \nu < 8;\;\nonumber\\
&&h_{9}=10-\nu,\;\;8 < \nu < 9;\;\;\;\;\;\;
h_{10}=\nu-8,\;\;\;\;\;\; 9 < \nu < 10;\nonumber\\
&&h_{11}=12-\nu,\;\;10 < \nu < 11;\;\;\;
h_{12}=\nu-10,\; 11 < \nu < 12;\\
&&h_{13}=14-\nu,\;\; 12 < \nu < 13;\;\;
h_{14}=\nu-12,\;\; 13< \nu < 14;\nonumber\\
&&h_{15}=16-\nu,\;\; 14 < \nu < 15;\;\;
h_{16}=\nu-14,\;\; 15 < \nu < 16;\nonumber\\
&&h_{17}=18-\nu,\;\; 16 < \nu < 17;\;\;
h_{18}=\nu-16,\;\; 17 < \nu < 18;\nonumber\\
&&h_{19}=20-\nu,\;\; 18 < \nu < 19;\;\;
h_{20}=\nu-18,\;\; 19 < \nu < 20;\nonumber\\
&&h_{21}=22-\nu,\;\; 20 < \nu < 21;\;\;
h_{22}=\nu-20,\;\; 21 < \nu < 22;\nonumber\\
&&h_{23}=24-\nu,\;\; 22 < \nu < 23;\;\;
h_{24}=\nu-22,\;\; 23 < \nu < 24;\nonumber\\
&&h_{25}=26-\nu,\;\; 24 < \nu < 25;\;\;
h_{26}=\nu-24,\;\; 25 < \nu < 26;\nonumber\\ 
&&etc.\nonumber
\end{eqnarray}

In this way, $h_{i}$, classifies the excitations in equivalence 
classes, whose elements are the statistics, $\nu$. On the 
other hand, we know that for path integration on multiply connected 
spaces\cite{R4}, we need to assign different weights, $\chi$, to 
homotopically disconnected paths. So, in terms of our scheme, 
as we just said, $h_{i}$, labels these classes, which form a group, 
the fundamental group $\Pi_{1}({\cal P})$, with elements, $\chi=\exp
\left\{\imath \nu \varphi\right\}$. We also have that, for the 
Abelian representation of the group, 
the weights satisfy the constraints\cite{R4}:

\begin{eqnarray}
&&|\chi\left\{\nu_{i}\right\}_{h_{i}}|=1,\\
&&\chi\left\{\nu_{i}\right\}_{h_{i}}\chi\left\{\nu_{j}
\right\}_{h_{j}}=
\chi\left\{\nu_{i}\circ\nu_{j}\right\}_{h_{i\circ j}}.\nonumber
\end{eqnarray}

\noindent The weights, $\chi$, represent a phase which 
we assign to the paths that contribute to the propagator:
 
\begin{eqnarray}
{\cal K}(q^{\prime},t^{\prime};q,t)=
\sum_{\left\{\nu\right\}_{h}}\chi
\left\{\nu\right\}_{h}{\cal K}^{\left\{\nu\right\}_{h}}
(q^{\prime},t^{\prime};q,t).
\end{eqnarray}
 
\noindent We observe that, for the composition law, 
for example, with $\nu_{1}+\nu_{2}$, we do not consider 
integers since $\nu$ can only be a rational number with 
odd denominator ( see again the intervals of $\nu$ ). On the 
other hand, we verify that some compositions with elements 
of the same class give another element either in or out 
of the class. The same occurs for the distinct classes 
( this can be, perhaps, a non-Abelian manifestation of the group ). 
But this does not matter, because we only consider paths into 
within the same class for the path integration. For example, 
consider the classes:
   
\begin{eqnarray}
&&\left\{\frac{2}{3},\frac{8}{3},\cdots
\right\}_{h=\frac{4}{3}};\;\;
\left\{\frac{3}{5},\frac{7}{5},\cdots
\right\}_{h=\frac{7}{5}};\\
&&\left\{\frac{5}{9},\frac{13}{9},\cdots
\right\}_{h=\frac{13}{9}};\;\;
\left\{\frac{4}{7},\frac{10}{7},\cdots
\right\}_{h=\frac{10}{7}}.\nonumber
\end{eqnarray}

\noindent If we compose, $\left\{\frac{2}{3}
\right\}_{h=\frac{4}{3}}$ with 
$\left\{\frac{3}{5}\right\}_{h=\frac{7}{5}}$
 we get $\left\{
\frac{19}{15}\right\}_{h=\frac{19}{15}}$; if
 we compose $\left
\{\frac{2}{3}\right\}_{h=\frac{4}{3}}$ with
 $\left\{\frac{8}{3}
\right\}_{h=\frac{4}{3}}$ we get $\left\{\frac{10}{3}
\right\}
_{h=\frac{4}{3}}$; if we compose $\left\{\frac{3}{5}
\right\}_{h=\frac{7}{5}}$ with $\left\{\frac{7}{5}
\right\}_{h=\frac{7}{5}}$ we get $\left\{ 2\right\}$
 which is not defined, 
etc. All of this, expresses the intricacy of
 the anyonic model for such a 
physical phenomenon as the FQHE.

Now, a point of theoretical interest: Under the exchange of 
two quasiholes\cite{R4} we have that the factor $
\left(z_{\alpha}-z_{\beta}\right)^{\nu}$ of the Laughlin 
function, with $\nu=\frac{1}{m}+2p_{1}$, where $m=3,5,7,
\cdots$ and $p_{1}$ is a positive integer, produces the 
condition on the phase:
 
\begin{equation}
\exp\left\{\imath\pi\left(\frac{1}{m}+2p_{1}\right)\right\}=
\exp\left\{\imath\pi\frac{1}{m}\right\}.
\end{equation}
 
\noindent Observe that, for $m=3$ and $p_{1}=1$, if we 
take into account our formulas for $h_{i}$, we get: 
$\left\{\frac{1}{3},\frac{7}{3}\right\}_{h=\frac{5}{3}}$; 
for $m=5$, $p_{1}=1$, we get $\left\{\frac{1}{5},
\frac{11}{5}\right\}_{h=\frac{9}{5}}$; for $m=7$,
$p_{1}=1$, we get $\left\{\frac{1}{7},\frac{15}{7}
\right\}_{h=\frac{13}{7}}$,etc. What does this tell us ? 
This simply confirms our hierarchy scheme for 
the filling factors: {\it the Hausdorff dimension 
classifies the collective excitations into equivalence 
classes and, in fact, the weights are elements of the Braid group }.

Another point, which we can touch on is, with respect to the largest 
charge gaps, these are observed for $\nu=\frac{1}{3}$ and $\nu=
\frac{2}{3}$\cite{R5}. Second in our scheme, $\left\{\frac{1}{3}
\right\}_{h=\frac{5}{3}}$ and $
\left\{\frac{2}{3}\right\}_{h=\frac{4}{3}}$, they are in different 
classes but the collective excitations for one case and other one 
is like a {\it quasi-bosonic regime}, $h\sim 2$, and 
{\it a quasi-fermionic regime}, $h\sim 1$, respectively. 
Thus, the interacting planar electron system presents this 
subtle behavior for all filling factors. Of course, this 
explains why some series of fractions are more favourable 
than others. Now, a last comment, which reinforces our 
classification of anyonic excitations, observe the mirror 
symmetry behind the construction of $h$.

\acknowledgments
I would like to thank Steven F. Durrant for reading 
the manuscript.

\end{document}